\documentclass[final,5p,twocolumn]{elsarticle}

\usepackage{graphicx}
\usepackage{tabularx}
\usepackage{dcolumn}
\usepackage{color}
\usepackage{nicefrac}
\usepackage{amsmath}
\usepackage{amssymb}
\usepackage{bm}
\usepackage{tabularx}
\usepackage{array}
\usepackage[T1]{fontenc}

\newcommand{\GOSIA}{\textsc{gosia}}

\newcommand{\efm}{\ensuremath{\mathrm{e}^2\mathrm{fm}^4}}
\newcommand{\mgcm}{\ensuremath{\mathrm{mg/cm}^{2}}}
\newcommand{\isot}[2]{\ensuremath{{}^{#1}\mathrm{#2}}}
\newcommand{\isotope}[4]{\ensuremath{{}^{#1}_{#3}\mathrm{#2}_{#4}}}

\journal{Physics Letters B}

\begin{document}

\begin{frontmatter}


\title{Suppressed Electric Quadrupole Collectivity in $^{49}$Ti}

\author[ornl,doe]{T.~J.~Gray}
\fntext[doe]{This manuscript has been authored by UT-Battelle, LLC, under contract DE-AC05-00OR22725 with the US Department of Energy (DOE). The US government retains and the publisher, by accepting the article for publication, acknowledges that the US government retains a nonexclusive, paid-up, irrevocable, worldwide license to publish or reproduce the published form of this manuscript, or allow others to do so, for US government purposes. DOE will provide public access to these results of federally sponsored research in accordance with the DOE Public Access Plan (http://energy.gov/downloads/doe-public-access-plan).}
\affiliation[doe]{
  organiztion={Physics Division, Oak Ridge National Laboratory},
  city={Oak Ridge},
  state={Tennessee},
  postcode={37831},
  coutnry={USA}
  }
\author[doe]{J.~M.~Allmond}
\author[fsu]{C.~Benetti}
\affiliation[fsu]{
  organiation={Department of Physics, Florida State University},
  city={Tallahassee},
  state={Florida},
  postcode={32306},
  country={USA}
}
\author[fsu]{C.~Wibisono}
\author[fsu]{L.~Baby}
\author[infn]{A.~Gargano}
\affiliation[infn]{
  organization={Istituto Nazionale di Fisica Nucleare, Complesso
    Universitario di Monte S. Angelo, Via Cintia},
  postcode={I-80126},
  city={Napoli},
  country={Italy}
  }
\author[tud,emmi,mpi]{T.~Miyagi}
\affiliation[tud]{
  organization={Department of Physics, Technische
    Universit{\"a}t Darmstadt},
  city={Darmstadt},
  country={Germany}
}
\affiliation[emmi]{
  organization={ExtreMe Matter Institute EMMI, GSI Helmholtzzentrum
    f\"ur Schwerionenforschung GmbH},
  postcode={64291},
  city={Darmstadt},
  country={Germany}
}
\affiliation[mpi]{
  organization={Max-Planck-Institut f\"ur Kernphysik, Saupfercheckweg
    1},
  postcode={69117},
  city={Heidelberg},
  country={Germany}
}
\author[ornl]{A.~O.~Macchiavelli}
\author[anu]{A.~E.~Stuchbery}
\affiliation[anu]{
  organization={Department of Nuclear Physics and Accelerator Applications, Research School of
    Physics, Australian National University},
  city = {Canberra},
  state={ACT},
  postcode={2601},
  country={Australia}
}
\author[git]{J.~L.~Wood}
\affiliation[git]{
  organization={School of Physics, Georgia Institute of Technology},
  city={Atlanta},
  state={Georgia},
  postcode={30332},
  country={USA}
}
\author[fsu]{S.~Ajayi}
\author[fsu]{J.~Aragon}
\author[utk]{B.~W.~Asher}
\affiliation[utk]{
  organization={Department of Physics and Astronomy, University of
    Tennessee},
  city={Knoxville},
  state={Tennessee},
  postcode={37966},
  country={USA}
}
\author[fsu]{P.~Barber}
\author[fsu]{S.~Bhattacharya}
\author[fsu]{R.~Boisseau}
\author[utk]{J.~M.~Christie}
\author[fsu]{A.~L.~Conley}
\author[fsu]{P.~De~Rosa}
\author[ornl]{D.~T.~Dowling}
\author[fsu]{C.~Esparza}
\author[fsu]{J.~Gibbons}
\author[fsu]{K.~Hanselman}
\author[triumf,mcgill]{J. D. Holt}
\affiliation[triumf]{
  organization={TRIUMF},
  address={4004 Wesbrook Mall},
  city={Vancouver},
  state={BC},
  postcode={V6T 2A3},
  country={Canada}
}
\affiliation[mcgill]{
  organization={Department of Physics, McGill University},
  city={Montr\'eal},
  state={QC},
  postcode={H3A 2T8},
  country={Canada}
}
\author[lsu]{S.~Lopez-Caceres}
\affiliation[lsu]{
  organization={Department of Physics and Astronomy, Louisiana State
    University},
  city={Baton Rouge},
  state={Louisiana},
  postcode={70803},
  state={USA}
}
\author[fsu]{E.~Lopez~Saavedra}
\author[fsu]{G.~W.~McCann}
\author[fsu]{A.~Morelock}
\author[fsu]{B.~Kelly}
\author[ornl]{T.~T.~King}
\author[ornl]{B.~C.~Rasco}
\author[fsu]{V.~Sitaraman}
\author[fsu]{S.~L.~Tabor}
\author[fsu]{E.~Temanson}
\author[fsu]{V.~Tripathi}
\author[fsu]{I.~Wiedenh\"over}
\author[scsu]{R.~B.~Yadav}
\affiliation[scsu]{
  organization={Department of Biological and Physical Sciences, South
    Carolina State University},
  city={Orangeburg},
  state={South Carolina},
  postcode={29117},
  country={USA}
}
\date{\today}

\begin{abstract}

Single-step Coulomb excitation of $^{46,48,49,50}$Ti is presented. A complete set of $E2$ matrix elements for the quintuplet of states in $^{49}$Ti, centered on the $2^+$ core excitation, was measured for the first time. A total of nine $E2$ matrix elements are reported, four of which were previously unknown. $^{49}_{22}$Ti$_{27}$ shows a $20\%$ quenching in electric quadrupole transition strength as compared to its semi-magic $^{50}_{22}$Ti$_{28}$ neighbour. This $20\%$ quenching, while empirically unprecedented, can be explained with a remarkably simple two-state mixing model, which is also consistent with other ground-state properties such as the magnetic dipole moment and electric quadrupole moment. A connection to nucleon transfer data and the quenching of single-particle strength is also demonstrated. The simplicity of the $^{49}$Ti-$^{50}$Ti pair (i.e., approximate single-$j$ $0f_{7/2}$ valence space and isolation of yrast states from non-yrast states) provides a unique opportunity to disentangle otherwise competing effects in the ground-state properties of atomic nuclei, the emergence of collectivity, and the role of proton-neutron interactions.

\end{abstract}

\end{frontmatter}

\par
Atomic nuclei are finite many body quantum systems that exhibit both single particle and collective degrees of freedom.
How nuclear collectivity and deformation emerge from the underlying single-particle motion of protons and neutrons has remained a priority open question for over a half century. The leading view is that correlations due to the long-range proton-neutron ($PN$) residual interaction drive collectivity and deformation~\cite{deShalit1953,Talmi1962,Federman1979} after they overcome the short-range proton-proton ($PP$) and neutron-neutron ($NN$) pairing correlations, which hold nuclei to spherical shapes. A natural consequence of this simple schematic idea is that an increase in the number of valence protons and/or neutrons outside some inert double-magic core should lead to more collective behaviour. This view is commonly expressed in the $N_p N_n$ scheme~\cite{Casten1985,Casten1985_2,Casten1987,Casten1996,Zhao2000}.

\par
One of the simplest approaches to investigating $PN$ interactions and emerging collectivity is to compare the $E2$ properties of a semi-magic ``core'' nucleus to its odd-$A$ neighbour, adjacent to the shell closure, i.e., the first step towards $PN$-driven collectivity.
If the $PN$ interactions are weak, the resulting system leads to the weak-coupling limit of the particle-core (PC) model by de-Shalit \cite{deShalit1961}. Within this limit, the odd-particle (or hole) couples to a presumed unperturbed core, leading to a degenerate multiplet of states, representing the different angular momentum coupling combinations, at each core excited state. In addition, the total summation of $B(E2{\uparrow})$ strength connected to the ground state is conserved. Within the shell model, this limit is reproduced when the $PN$ interactions are set to zero. The weak-coupling limit can be expected to work best when the separation of single-particle energies is large compared to the core excitation energies~\cite{deShalit1961,Braunstein1962,Gove1963}. 

\par
Curiously, \isot{63,65}{Cu}, \isot{93}{Mo}, and \isot{113,115}{In} show no change in $E2$ excitation strength as compared to their semi-magic cores, cf.,~Fig.~4 in Ref.~\cite{Gray2020} and Ref.~\cite{Tuttle1976}, despite showing complexity beyond the weak-coupling limit 
(see also Fig.~\ref{fig:global_comp} below). A significant $40\%$ increase in $E2$ strength was recently found in \isot{129}{Sb} relative to semi-magic \isot{128}{Sn}~\cite{Gray2020}. This was empirically correlated to the small $B(E2)$ strength of the \isot{128}{Sn} core and proximity to the double-magic nucleus \isot{132}{Sn}. The enhanced collectivity was interpreted within the shell model as being due to constructive quadrupole coherence in the wave functions stemming from the proton-neutron residual interactions mixing the valence core neutrons in a multi-$j$ space, predominantly $\nu s_{1/2}$, $\nu d_{3/2}$, and $\nu h_{11/2}$. 


\par
In this Letter, we present single-step Coulomb excitation results of \isot{46,48,49,50}{Ti}. 
Contrary to empirical observations highlighted in Fig.~4 of Ref.~\cite{Gray2020}, we find a suppression of electric quadrupole strength in \isotope{49}{Ti}{22}{27} with respect to its semi-magic neighbour, \isotope{50}{Ti}{22}{28}, with only two protons outside the doubly closed \isotope{48}{Ca}{20}{28}. Thus, by increasing $N_p\times N_n$ from 0 to 2, the quadrupole transition strength decreases. Of all the cases with $Z>8$, where both the semi-magic core and odd-$A$ neighbour are stable, \isot{50}{Ti} has the lowest $B(E2;0^{+} \rightarrow 2^{+})$ ~\cite{Pritychenko2016}, and, similar to \isot{128}{Sn}, it is near a double-shell closure, i.e., \isot{48}{Ca}. 

\par
The experiments were performed at the Florida State University (FSU) John D. Fox Laboratory, using the 9-MV tandem accelerator to deliver beams of \isot{46,48,49,50}{Ti}. The nuclei were incident on natural C and Al targets, 0.60- and 0.41-$\mgcm$ thick, respectively. The beam energies ranged between $74.0$ and $85.0$~MeV, with the safe criterion of 5~fm of separation maintained for each beam-target combination.
The CLARION2-TRINITY array was used to detect $\gamma$ rays and charged particles~\cite{CLARION2}. A total of nine Compton-suppressed HPGe Clover detectors were present in the array.
Two rings of the TRINITY particle detector were used to detect the lighter recoiling target nuclei, covering the angles $14^{\circ}-24^{\circ}$ and $34^{\circ}-44^{\circ}$. In addition, the array consisted of a zero-degree detector for stopping-power and beam-composition measurements.
The beam composition was measured approximately every $8$~hrs and contaminants due to the use of double stripping were consistently found to be $\leq 5\%$. The average beam purities for \isot{46,48,49,50}{Ti} were 99$\%$, 95$\%$, 95$\%$, and 98$\%$, respectively. For further details of the experimental setup, including stopping-power and beam-composition measurements,
see Ref.~\cite{CLARION2}.

\par
\begin{figure}[t]
    \centering
       \includegraphics[width=\columnwidth]{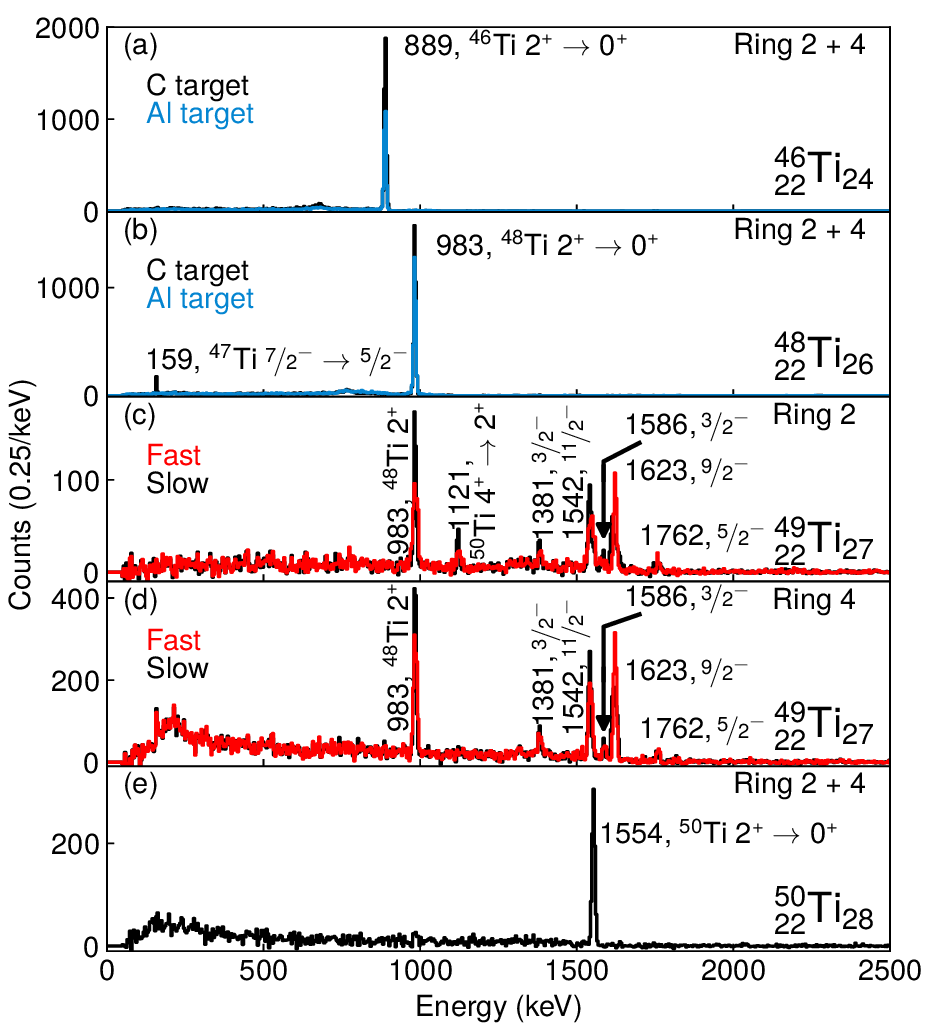}
       \caption{Doppler-corrected $\gamma$-ray spectra from Coulomb excitation of \isot{46,48,49,50}{Ti}. (a--b) Carbon and aluminium targets were used for \isot{46,48}{Ti}, which are shown in black and blue spectra, respectively. The 159-keV peak is due to \isot{47}{Ti} contamination in the beam. (c--d) \isot{49}{Ti} spectra with ``slow'' Doppler-correction (black), and ``fast'' Doppler-correction (red). (e) \isot{50}{Ti} spectrum.}
    \label{fig:espectra}
\end{figure}

Doppler-corrected $\gamma$-ray spectra for \isot{46,48,49,50}{Ti} are shown in Fig.~\ref{fig:espectra}. Both C and Al targets were used for \isot{46,48}{Ti} experiments, while only the C target was used for \isot{49,50}{Ti}. The prominent peaks correspond to $2_1^+$ excitations. \isot{49}{Ti} has two transitions which decay primarily through $M1$ decay, with much shorter lifetimes than the other states. Thus, two sets of Doppler-correction $v/c$ values were used, one for the ``slow'' $E2$ transitions (black), and one for the ``fast'' $M1$ transitions (red). The mean $v/c$ values per particle-detector ring were determined empirically by fitting the observed $\gamma$-ray energies as a function of the cosine between the $\gamma$ ray and particle to minimize the FWHM of the Doppler-corrected peaks. The measured values were consistent with the two, ``fast'' and ``slow'', asymptotic limits of $v/c$, where the ``fast'' transitions decay immediately within the target and the ``slow'' transitions decay after exiting the target. The 1121-keV peak in Fig.~\ref{fig:espectra}(c) is due to the $4^{+} \rightarrow 2^{+}$ transition in \isot{50}{Ti}. This comes from the $\isot{49}{Ti}(\isot{13}{C},\isot{12}{C})\isot{50}{Ti}$ sub-barrier reaction from the \isot{\mathrm{nat}}{C} target. The reaction populates a number of high-lying states which all decay through the $4^{+} \rightarrow 2^{+} \rightarrow 0^{+}$ cascade, as verified in a subsequent measurement with a \isot{13}{C} target. Thus, the 1121-keV $4^{+} \rightarrow 2^+$ and 1554-keV $2^+ \rightarrow 0^+$ transitions are present in the Ring 2 spectrum in equal intensity. The 1554-keV transition is only partially resolved from the 1542-keV transition in \isot{49}{Ti}, and to account for this the efficiency-corrected 1121-keV area was subtracted from the combined 1542/1554-keV peak area for the \GOSIA~analysis.


\par
Matrix elements were extracted from the efficiency-corrected particle-$\gamma$ peak areas, normalized to ``Rutherford'' particle singles counts (i.e., particle-$\gamma$/particle $\propto \sigma_{\mathrm{Coulex}}/\sigma_{\mathrm{Rutherford}}$) on a ring-by-ring basis, with the semi-classical Coulomb excitation program \GOSIA~\cite{GOSIA}. The energy loss through the target was measured using the downstream zero-degree detector as 19.5(5)~MeV for the C target and 6.7(5)~MeV for the Al target. These energy losses were consistent across the face of each target foil. The analysis procedures, including necessary corrections, were similar to Refs.~\cite{Allmond2011,Allmond2013,Allmond2014,Allmond2015,Allmond2017,Stuchbery2013,Gray2020}. Systematic uncertainties due to unknown branches, unknown $\delta(E2/M1)$ mixing ratios, beam composition, H-scattering from surface contamination on the target, known and unknown quadrupole moments, absolute $\gamma$-ray efficiency, energy loss through the target, and particle-detector angular offsets were accounted for. The extracted $B(E2)$ values are given in Table~\ref{tab:be2s}.

\begin{table*}[t]
    \centering
    \caption{$B(E2{\uparrow})$ values extracted in the present work are given with statistical and systematic uncertainties in the second set of parenthesis. See text for details on the theoretical calculations, which span the zeroth order particle-core coupling with no interaction to the most recent state-of-the-art \textit{ab initio} interactions. Ratios of the \isot{49}{Ti} $\Sigma B(E2{\uparrow})$ to \isot{50}{Ti} are provided.}

    
    
   {\small
   \setlength{\tabcolsep}{3pt}
\begin{tabular}{>{\raggedleft}p{1.2cm}|>{\centering}p{0.8cm}>{\centering}p{0.7cm}>{\centering}p{0.7cm}>{\centering}p{1.4cm}>{\centering}p{2.65cm}>{\centering}p{1.25cm}>{\centering}p{1.25cm}>{\centering}p{1.2cm}>{\centering}p{1.2cm}>{\centering}p{1.7cm}>{\centering\arraybackslash}p{1.7cm}}
\hline \hline
   &&&&&&&&&&perturbative& \\
   &&&&&&$V_{QQ}=0$&$V_{QQ}\neq0$&MBZ&KB3&\textit{ab initio}& \textit{ab initio}\\
   
                      & $E_f$    & $J^{\pi}_i$ & $J^{\pi}_f$  & $B(E2{\uparrow})$ & Literature $B(E2{\uparrow})$                             & PC & PC-QQ       & SM-$f_{7/2}$ & SM-$fp$ & SM-$fpg_{9/2}$  & VS-IMSRG   \\ 
                      & (keV)    &             &           &  (\efm)    &  (\efm)      &  (\efm)       &  (\efm)                                                  & (\efm)    & (\efm)                & (\efm)                & (\efm)  \\ \hline
                      
        \isot{46}{Ti} &  889     &    $0^{+}$  &     $2^{+}$  &  940(10)(50)      & 
        951(25)~\cite{Pritychenko2016} &       && 405   &      519          & 454                   & 489     \\ \hline
        \isot{48}{Ti} &  983     &   $0^{+}$   &    $2^{+}$   &  700(10)(40)      & 
        662(29)~\cite{Pritychenko2016} &       && 364  &      426           & 351                   & 375     \\ \hline  
        \isot{49}{Ti} &  1381    &  $7/2^{-}$  &    $3/2^{-}$ &  12.0(12)(8)      & 16.8(22)~\cite{Mando1981}                                &    0   &  0                   & 0 &11 & 0                       & 0.34        \\
                      &   1542   &   $7/2^{-}$ &   $11/2^{-}$ &  81(4)(6)         & 97(10)~\cite{Mando1981}                                 & 106 & 86    & 125                & 82  & 75                      & 80        \\
                      &   1586   &   $7/2^{-}$ &    $3/2^{-}$ &  17.0(21)(12)     &                                                          & 35 & 31    & 54                    &22 & 27                      & 30        \\  
                      &   1623   &   $7/2^{-}$ &   $9/2^{-}$  &  135(5)(9)        &                                                          & 88 & 87    & 64                    & 102 & 102                      & 111        \\
                      &   1762   &   $7/2^{-}$ &   $5/2^{-}$  &  32(5)(2)         &                                                          & 53 & 27    & 43                    &30 & 37                      & 38        \\
                      &   (2262) & $7/2^{-}$   &  $7/2^{-}$   & $0(^{+28}_{-0})$             &                                                          & 70 & 12   & 18                    &14 & 32                      & 28        \\
                                            & & & & & & & & & & & \\

       $\Sigma B(E2{\uparrow})$           &          &             &              &  $276(^{+29}_{-8})(19)$      &                                                         & 352 & 243   & 303                 & 261 & 273                      & 287        \\
       \textbf{Ratio}: & & & & $\bm{0.78(^{+9}_{-4})}$ & & \textbf{1} & \textbf{0.69} & \textbf{0.85} & \textbf{0.79} & \textbf{0.80} & \textbf{0.77}\\
       \hline
       \isot{50}{Ti}  &   1554   &    $0^{+}$  &   $2^{+}$    &  352(17)(23)      & 
       275(16)~\cite{Pritychenko2016} &      & & 355                   & 330 & 340                   & 371     \\  \hline \hline

    \end{tabular}}

    \label{tab:be2s}
\end{table*}

\par
Overall, the current results compare well to previous measurements of $B(E2)$ strengths in the Ti chain.
There is good agreement for \isot{46,48}{Ti} but our measurement is higher than the most recent evaluation \cite{Pritychenko2016} for \isot{50}{Ti}. The present measurement is the first example of safe Coulomb excitation of \isot{50}{Ti} in inverse-kinematics. There has been only one previous safe-Coulex measurement~\cite{Towsley1975}, which is consistent with the present measurement; other measurements are either from unsafe Coulex~\cite{Simpson1965, Hausser1970, Vasilev1962, Afonin1968} or lifetime measurements using the Doppler-shift attenuation method~\cite{Ward1972, Speidel2000, Speidel2003}. The lifetime measurements from transient field $g$-factor experiments \cite{Speidel2000, Speidel2003} display the same systematic deviation as recently demonstrated for the \textrm{Ni} and \textrm{Sn} isotopes ~\cite{Allmond2014, Allmond2015}.

\begin{figure}[t]
    \centering
    \includegraphics[width=\columnwidth]{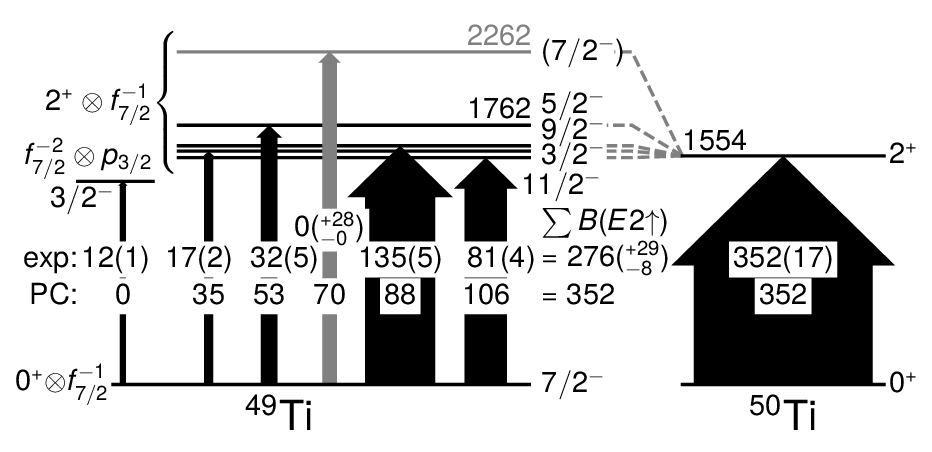}
    \caption{Low-lying level schemes of \isot{49,50}{Ti}, with $B(E2)$ excitation strengths given in \efm. The total quadrupole excitation strength in \isot{50}{Ti} is fragmented between multiplet members in \isot{49}{Ti} and the total strength is reduced by $\approx 20\%$. ``PC'' refers to the simple weak-coupling limit of the particle-core coupling model of de-Shalit \cite{deShalit1961} with zero interactions, which fragments but does not suppress or enhance the quadrupole strength.}
    
    \label{fig:levelscheme}
\end{figure}

\par
Complete single-step Coulomb excitation of \isot{49}{Ti} is presented for the first time here. Figure~\ref{fig:levelscheme} shows the low-lying level scheme, along with the extracted $B(E2)$ strengths. The $(2J + 1)$-weighted energy sum of the four states between $1542$ and $1762$~keV is 1614~keV, close to the $2^{+}$ energy of $1554$~keV in \isot{50}{Ti}. We observe population of four $2^{+} \otimes f_{7/2}$ multiplet states with $J^{\pi} = 3/2^-,5/2^-,9/2^-,11/2^-$, as well as a $3/2^-$ ``intruder'' state with an expected $\nu f_{7/2}^{-2} p_{3/2}$ configuration. The only previous measurements of $B(E2)$ strengths in \isot{49}{Ti} are from Coulomb excitation~\cite{Mando1981}, which give slightly larger values for the two lowest excited states. The $7/2^-$ multiplet member has not been identified experimentally but a candidate exists at 2262~keV. No peak at 2262 keV was observed, but the background was used to establish a one-sigma upper uncertainty, $B(E2; 7/2_1^- \rightarrow 7/2_2^-) = 0(^{+28}_{-0})$~e$^{2}$fm$^{2}$.
Within the simple weak-coupling limit of the particle-core coupling (PC) model by de-Shalit \cite{deShalit1961}, the sum of the $B(E2)$ transition strengths connected to the ground state should be identical between the particle-core and core systems, see Refs.~\cite{deShalit1961,Gray2020}. Shell-model calculations reproduce this limit if the $PN$ interactions are set to zero, and thus deviations provide an important way of investigating the role of $PN$ interactions in the development of $E2$ collectivity.
As can be seen in Fig.~\ref{fig:levelscheme}, some experimental transitions are enhanced and some are suppressed with respect to the individual weak-coupling PC values.

\par
The sum of the $B(E2{\uparrow})$ strengths for the five transitions measured, including uncertainty associated with the unobserved $7/2^-$ member, is $276(^{+29}_{-8})(19)$~\efm, where the first and second set of parentheses give the statistical and systematic uncertainties, respectively. This sum is smaller than the $B(E2; 0^+ \rightarrow 2^+)$ in \isot{50}{Ti} of $352(17)(23)$~\efm, giving a ratio of $0.78(^{+9}_{-4})$, where systematic effects largely cancel between the \isot{49}{Ti} and \isot{50}{Ti} measurements. Thus, we find $\Sigma B(E2{\uparrow})$ in \isot{49}{Ti} to be suppressed by roughly $20\%$ with respect to the semi-magic \isot{50}{Ti} core.

\begin{figure}[t]
    \centering
    \includegraphics[width=\columnwidth]{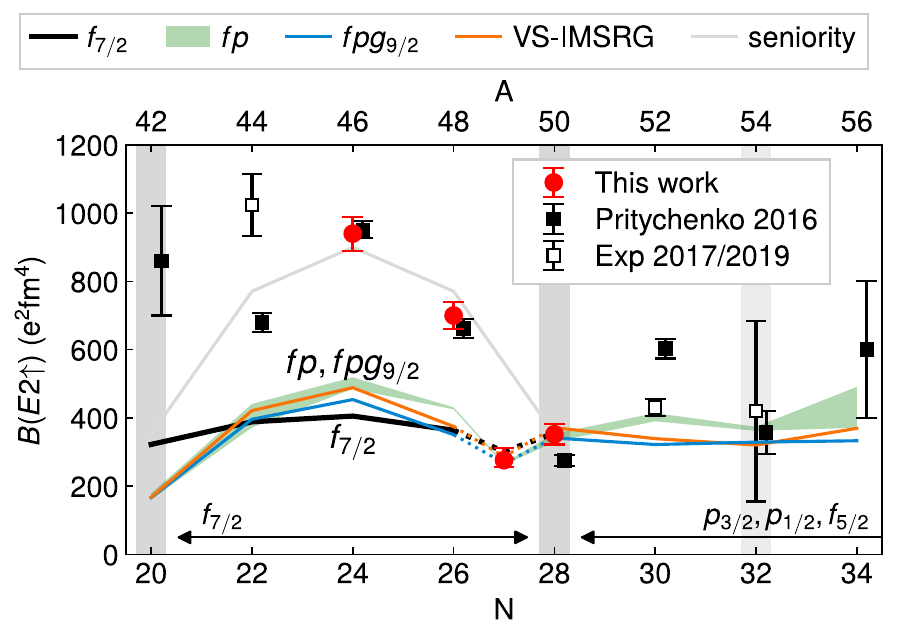}
    \caption{$B(E2;0^{+} \rightarrow 2^{+})$ systematics across the \textrm{Ti} ($Z=22$) isotopic chain~\cite{Pritychenko2016,Arnswald2017, Goldkuhle2019}. A phenomenological seniority curve, fitted to \isot{42}{Ca} and \isot{50}{Ti}, is shown to guide the eye~\cite{Talmi1971,Talmi1993,Morales2011,Macchiavelli2014}. See text for additional theoretical details.} 
    
    
    \label{fig:systematics}
\end{figure}

\par
In order to understand the origin of the quenched electric quadrupole strength in \isot{49}{Ti}, we first explore the $B(E2{\uparrow})$ systematics of the \textrm{Ti} ($Z=22$) chain, see Fig.~\ref{fig:systematics}. Shell-model calculations were performed in the restricted $f_{7/2}$ space (black curve) with the MBZ interaction \cite{McCullen1964} and in the larger $fp$ space with the KB3 \cite{Poves1981} and GXPF1 \cite{Honma2002} interactions, including the monopole-adjusted KB3G \cite{Poves2001} and GXPF1A \cite{Honma2005} interactions. The KB3, KB3G, GXPF1, and GXPF1A interactions yield similar $B(E2{\uparrow})$ results and are represented as a shaded green curve in Fig.~\ref{fig:systematics}. In addition, calculations were performed with non-empirical interactions derived from realistic nuclear forces by means of \textit{ab initio} many-body methods. Results of two calculations are presented. In the first (blue curve), the effective shell-model Hamiltonian is constructed within the framework of the many-body perturbation approach starting from chiral nucleon-nucleon~(NN) plus three-nucleon~(3N) potentials~\cite{Coraggio2020,Coraggio2021}, and is defined in the $fpg_{9/2}$ valence space. The second set of interactions (orange curve) employ a non-perturbative approach. They are based on the valence-space formulation of the in-medium similarity renormalization group (VS-IMSRG)~\cite{Stroberg2019} with recent developments~\cite{Miyagi2022,Stroberg2017} using the EM 1.8/2.0 interaction~\cite{Hebeler2011,Entem2003}. The NN and 3N matrix elements were computed with the \texttt{NuHamil} code~\cite{Miyagi2023}, and the VS-IMSRG step was performed with the \texttt{imsrg++} code by S.~R.~Stroberg~\cite{imsrg}. The diagonalization was carried out in the $fp$ space. The KSHELL diagonalization program~\cite{KSHELL} was used in all the calculations. 



\par
Effective charges of $e_p = 1.7$ and $e_n = 0.5$ were adopted for the $f_{7/2}$ space, set to reproduce the $B(E2; 0^+ \rightarrow 2^+)$ of \isot{50}{Ti}. Effective charges of $e_p = 1.1$ and $e_n = 0.6$ were used for the $fp$ and $fpg_{9/2}$ spaces, which were set to reproduce the $B(E2; 0^+ \rightarrow 2^+)$ strengths in \isot{50}{Ti} and \isot{48}{Ca}; these effective charges are similar to those adopted for describing the $Z$ and $N=28$ chains \cite{Allmond2014}. As shown in Fig.~\ref{fig:systematics}, the larger valence spaces require smaller effective charges while also generating larger $E2$ strength near midshell. All calculations reproduce the data for \isot{49}{Ti} relative to \isot{50}{Ti}, where the shell-model calculations are expected to work best; this $E2$ ratio is insensitive to the effective charge values. However, they all fail to reproduce the large increase in $B(E2{\uparrow})$ from \isot{49}{Ti} to \isot{48}{Ti}. This failure may be a result of the limited basis space; excitations across $N$ and $Z=20$ are not included and are known to be relevant for the region, particularly closer to $N=Z$; see discussions on the Ca isotopes in Refs.~\cite{SDPF-Mix-40Ca, SDPF-Mix-42Ca, Ideguchi2022, Stuchbery2022}. Unfortunately, the basis space for such expanded calculations is currently too large to be practical for the stable Ti isotopes. See also discussions on quenching of $E2$ strength within \textit{ab initio} calculations due to the difficulty in capturing highly collective degrees of freedom~\cite{Hend18E2,Stro22E2}. Clearly, both the number of active nucleons and the number of configurations available (including the number of oscillator shells and coupling to the giant quadrupole resonance) critically influence the generation of coherent sums of $E2$ strength. Nevertheless, excitations across $Z=20$ are less relevant for nuclei adjacent to \isot{48}{Ca}, which accounts for the good description of \isot{49,50}{Ti}. 


\begin{figure}[t]
    \centering
    \includegraphics[width=\columnwidth]{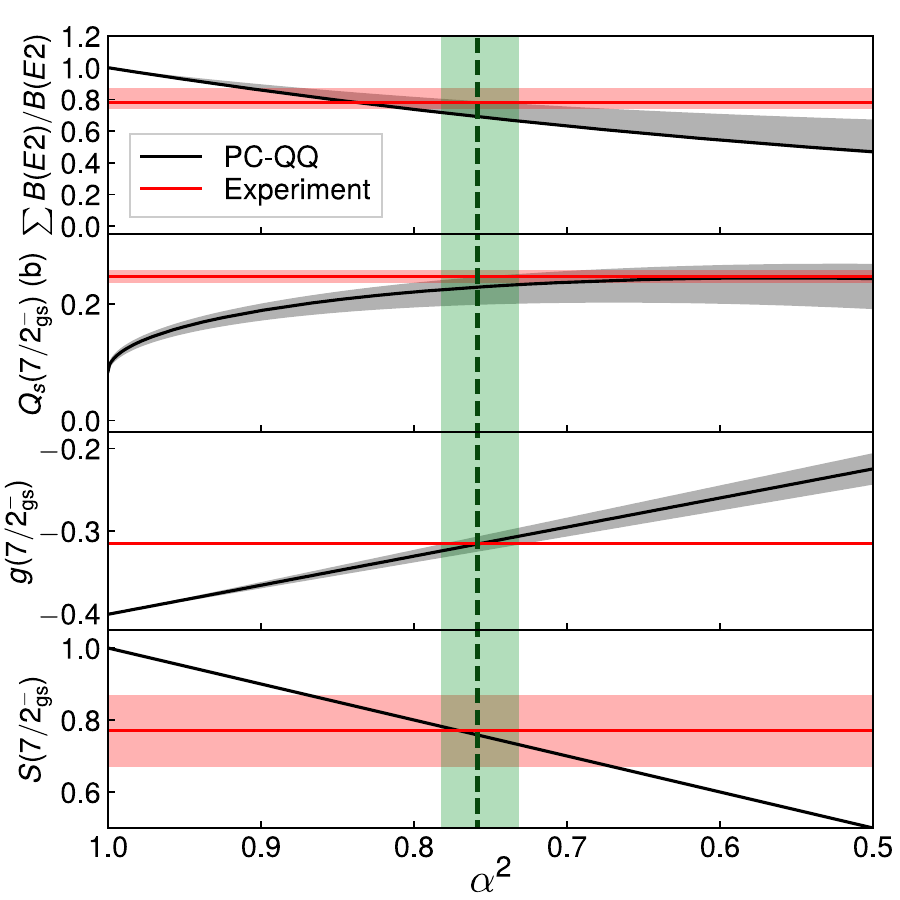}
    \caption{The PC-QQ model fitted to the electromagnetic-moment \cite{Jeffries1953, Drain1965, Bieron1999} and spectroscopic-factor \cite{Wilhjelm1968, Andersen1969, Fortier1978} data on \isot{49}{Ti}. The data are well correlated to $\alpha^2 = 0.76(3)$. The grey shaded regions correspond to experimental uncertainties in the empirical parameters used in the PC-QQ model. Thus, the $7/2_1^-$ ground state is $76\%$ $0^+\otimes f_{7/2}$ and $24\%$ $2^+\otimes f_{7/2}$}.

    
    \label{fig:twostate}
\end{figure}

\par
The suppressed $E2$ strength in \isot{49}{Ti} relative to semi-magic \isot{50}{Ti} appears to be a universal feature of all the calculations, including those in the truncated $f_{7/2}$ space. The underlying mechanism leading to this effect can be understood within a simple extension of the PC weak-coupling model of de-Shalit~\cite{deShalit1961} by including a finite quadrupole-quadrupole (QQ) interaction between the particle and core, as developed by Thankappan and True~\cite{Thankappan1965}. The $QQ$ interaction is a proxy for the $PN$ interaction. If the core is restricted to the $0^+$ ground state and $2^+$ excited state, the model reduces to simple two-state mixing between the $[0^+\otimes f_{7/2}]_{7/2^-}$ and $[2^+\otimes f_{7/2}]_{7/2^-}$ configurations with a mixing amplitude of $\alpha$. This naturally drives the $7/2^-$ member of the $2^+$ multiplet to high energy and it perturbs all $B(E2; 7/2_1^-\rightarrow J^-)$ values. Beyond the free parameter $\alpha$, the model presumes fixed empirical parameters for the odd-particle (or hole) and core matrix elements, namely the experimental electromagnetic moments of the \isot{47}{Ca} $7/2^-$ ground state~\cite{Garcia2015} and \isot{50}{Ti} $2^+$ excited state~\cite{Hausser1970, Towsley1975, Speidel2000}, respectively. The results of this simple PC-QQ model are shown in Fig.~\ref{fig:twostate}. The \isot{49}{Ti} data~\cite{Jeffries1953, Drain1965, Bieron1999, Wilhjelm1968, Andersen1969, Fortier1978} are well reproduced with a best-fit $\alpha^2 = 0.76(3)$, i.e., the $7/2_1$ ground state is $76\%$ $0^+\otimes f_{7/2}$ and $24\%$ $2^+\otimes f_{7/2}$. This is consistent with the SM-$fpg_{9/2}$ ($78\%$) and VS-IMSRG ($73\%$). We note that the low-lying $\nu p_{3/2}$ state is not present in the simple PC-QQ model. However, the introduction of this state would not perturb the total excitation strength but rather fragment it between multiple excited $3/2^-$ states, and thus a comparison of the experimental and PC-QQ $B(E2)$ sums is justified.






\par
A comparison between the experimental and calculated $B(E2{\uparrow})$ values for \isot{49}{Ti} are provided in Table~\ref{tab:be2s}, ranging from the zeroth order PC model to the most state-of-the-art \textit{ab initio} theories. While all but the zeroth-order PC model can reproduce suppressed $\Sigma B(E2{\uparrow})$ strength, the fragmentation of the individual $B(E2{\uparrow})$ transitions are in fact sensitive to the model space and $PN$ interactions. The shell-model calculations in the limited $f_{7/2}$ space with the empirical MBZ interaction perform the worst, followed by the simple PC-QQ model. We present only the KB3 results of the four empirically adjusted $fp$-space interactions (i.e., KB3, KB3G, GXPF1, and GXPF1A). Curiously, the KB3 and GXPF1 interactions in the $fp$ space give similar results and both have finite $7/2_1\rightarrow3/2_1$ strength to the 1p-2h intruder, which vanishes with the monopole-corrected KB3G and GXPF1A interactions. In addition, the KB3G and GXPF1A interactions give relative $B(E2{\uparrow})$ values that are very similar to the \textit{ab initio} interactions, which are frequently used to guide neutron-rich studies and potential sub-shell closures near $N=32,34$ and beyond.

\begin{figure}[t]
    \centering
    \includegraphics[width=\columnwidth]{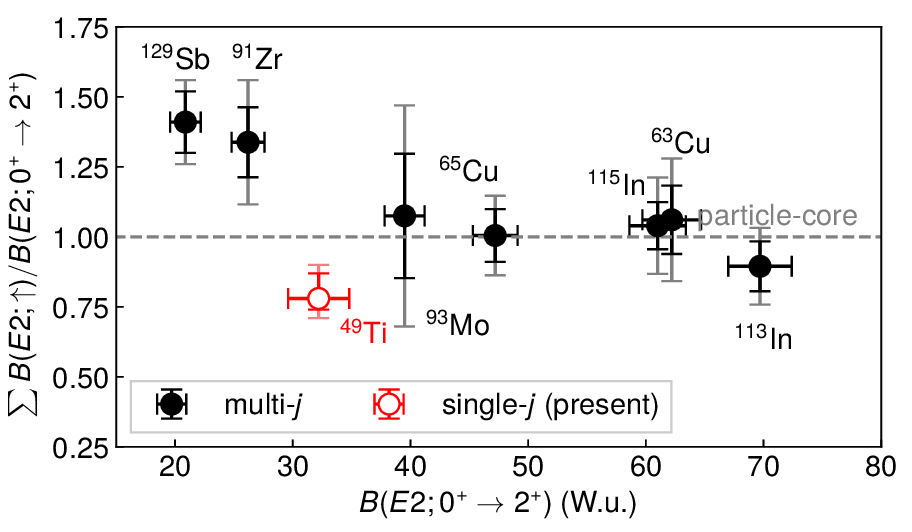}
    \caption{Global comparison of the particle-core sum rule across the isotopic chart. While an enhancement was found for \isot{129}{Sb} \cite{Gray2020}, a deficit was found for \isot{49}{Ti}. For cases with more collective semi-magic cores, the sum rule holds surprisingly well.}
    \label{fig:global_comp}
\end{figure}

\par
A global view of particle-core to core $\Sigma B(E2{\uparrow})$ ratios for nuclei adjacent to shell closures with unequivocal data is provided in Fig.~\ref{fig:global_comp}. Unlike the recent enhancement observed in \isot{129}{Sb}~\cite{Gray2020}, a deficit is found in \isot{49}{Ti}. One distinguishing characteristic of \isot{49}{Ti} is that, unlike the other cases, it possesses an approximate single-$j$ core. The two active protons in the $f_{7/2}$ space of \isot{50}{Ti} can only lead to four states, $0^+$, $2^+$, $4^+$, and $6^+$. The single-$j$ nature is empirically reflected in the fact that the first non-yrast state in \isot{50}{Ti} is above $6_1^+$, near 3.2~MeV~\cite{ENSDF}. The absence of low-lying non-yrast states in the core leads to deficits in the odd-mass $B(E2{\uparrow})$ strength. For example, the simple PC-QQ model with finite mixing gives deficits in the absence of non-yrast core states, even when $4_1^+$ and $6_1^+$ states are included (which otherwise have a negligible effect on the present results). 
This ``yrast isolation'' is unique to the \isot{49}{Ti}-\isot{50}{Ti} pair as compared to the other cases in Fig.~\ref{fig:global_comp}. It remains unclear why the cases with more collective semi-magic cores, typically near midshell, exhibit unity; one can speculate that the $PN$ contributions to the total $E2$ strength are negligibly small compared to the $PP$ and $NN$ contributions.

\par
To summarize, the first results from the CLARION2-TRINITY array --- Coulomb excitation of \isot{46,48,49,50}{Ti} --- have been presented. A complete set of $E2$ matrix elements for the quintuplet of states in $^{49}$Ti, centered on the $2^+$ core excitation, was measured for the first time. A quenching of the total electric quadrupole transition strength in $^{49}_{22}$Ti$_{27}$ by 20$\%$ is observed relative to semi-magic $^{50}_{22}$Ti$_{28}$, opposite to previous empirical observations involving multi-$j$ valence space nuclei. The anomalous trend is suggested to be primarily from the mixing of $[0^{+} \otimes \nu f_{7/2}^{-1}]_{7/2^-}$ and $[2^{+} \otimes \nu f_{7/2}^{-1}]_{7/2^-}$ configurations, and the relative isolation of the valence nucleons to the single-$j$ $0f_{7/2}$ shell. The $E2$ fragmentation pattern was shown to be very sensitive to the finer details of the valence space size, effective single-particle energies (or monopole corrections), and underlying $PN$ interactions, with implications extending to the neutron-rich $N=32,34$ region. Finally, the new results provide further evidence of the fundamental importance of configuration mixing (and the number of available configurations) in driving the emergence of $E2$ collectivity.




We would like to acknowledge the Center for Accelerator Target Science (CATS) and Matt Gott for making the C and Al foils used in this study and Alfredo Poves for useful discussions on the Shell-Model calculations. This material is based upon work supported in part by the U.S. Department of Energy, Office of Science, Office of Nuclear Physics under Contract No. DE-AC05-00OR22725 (ORNL). This work was also supported by the U.S. National Science Foundation under Grant No.~PHY-2012522 (FSU) and the Australian Research Council under grant No.~DP210101201. In addition, this work was supported in part by the Deutsche Forschungsgemeinschaft (DFG, German Research Foundation) --- Project-ID 279384907 --- SFB 1245, and the European Research Council (ERC) under the European Union’s Horizon 2020 research and innovation programme (Grant Agreement No. 101020842). The VS-IMSRG calculations were supported by NSERC under grants SAPIN-2018-00027 and RGPAS-2018-522453 and the Arthur B. McDonald Canadian Astroparticle Physics Research Institute. Calculations were performed with an allocation of computing resources at the J\"ulich Supercomputing Center and Cedar at WestGrid with The Digital Research Alliance of Canada. The publisher acknowledges the US government license to provide public access under the DOE Public Access Plan (http://energy.gov/downloads/doe-public-access-plan). 

\bibliography{Ti}

\end{document}